\documentclass[11pt]{article}
\usepackage{amssymb}
\usepackage{psfig}
\usepackage{graphicx}
\usepackage{mathrsfs}

\input tcilatex
\renewcommand{\theequation}{\thesection.\arabic{equation}}

\begin{document}

\title{Optimal Redeeming Strategy of Stock Loans}
\date{}
\author{Min Dai\thanks{%
Department of Mathematics, National University of Singapore (NUS),
Singapore. Dai is also an affiliated member of Risk Management
Institute, NUS. Partially supported by Singapore MOE AcRF grant
(No. R-146-000-096-112) and NUS RMI grant (No. R-703-000-004-720/646 and R-146-000-117-720/646).} \and %
Zuo Quan Xu\thanks{%
Mathematical Institute, University of Oxford, 24--29 St Giles,
Oxford OX1 3LB.}
} \maketitle

\begin{abstract}
A stock loan is a loan, secured by a stock, which gives the borrower the
right to redeem the stock at any time before or on the loan maturity. The
way of dividends distribution has a significant effect on the pricing of the
stock loan and the optimal redeeming strategy adopted by the borrower. We
present the pricing models subject to various ways of dividend distribution.
Since closed-form price formulas are generally not available, we provide a
thorough analysis to examine the optimal redeeming strategy. Numerical
results are presented as well.
\end{abstract}


\section{Introduction}

Optimal stopping problems in finance have gained growing interests
due to their close linkage with various optimal strategies. A
typical and well-known example is the American vanilla option
pricing model which has been been extensively studied in the
existing literature. Many researchers have also considered plenty of
more sophisticated models. For example, Gerber and Shiu (1996) and
Broadie and Detemple (1997) analyzed the optimal exercise strategy
for American options on multi-assets, which was further addressed by
Villeneuve (1999), Jiang (2002), Detemple et al. (2003), etc. Cheuk
and Vorst (1997), Windcliff et al. (2001), and Dai et al. (2004)
investigated the optimal shouting strategy for shout options. Dai
and Kwok (2006) characterized the optimal exercise strategy for
American-style Asian options and lookback options. Hu and Oksendal
(1998) studied the optimal strategy of an investment problem. Other
studies along this line include the pricing of game options, swing
options and convertible bonds, and the multiple-stopping problems in
finance, see Kifer (2000), Carmona and Touzi (2003), IBanez (2004),
Meinshausen and Hambly (2004), Dai and Kwok (2005), Dai and Kwok
(2008), etc.

In this paper, we take into consideration another optimal stopping problem
arising from the pricing of a financial product: \textit{stock loan, which }%
is a contract between a client (borrower) and a bank (lender). The borrower,
who owns one share of a stock, obtains a loan, from the lender with the
share as collateral. The borrower may redeem the stock at any time before or
on the loan maturity by repaying the lender the principal and a
predetermined load interest rate, or surrender the stock instead of repaying
the loan. The accumulative dividends may be gained by the borrower or the
lender, subject to the provision of the loan.

A natural pricing problem arises for both the borrower and the
lender: given the principal $K$ and the loan interest rate $\gamma
,$ what is the fair fee charged by the lender (referred to as the
price of the stock loan)? Moreover, the borrower is facing another
problem: what is the optimal redeeming strategy? Xia and Zhou (2007)
initiated the study of the above problems under the Black-Scholes
framework. Assuming that the loan is of infinite maturity and the
dividends accrued are gained by the lender until the borrower
redeems the stock, they revealed that the stock loan is essentially
an American call option with a possibly negative interest rate.
Moreover, they obtained a closed form price formula as well as the
analytical optimal redeeming strategy. Zhang and Zhou (2009) further
extended it to a regime switching market.

In this paper, we are concerned with finite maturity stock loans with
various ways of dividend distribution. Except for special cases, closed form
price formulas are no longer available. As in Xia and Zhou (2007), the
pricing model of a stock loan resembles that of (finite maturity) American
vanilla options if the dividends are gained by the lender before redemption.
However, other ways of dividends distribution may significantly alter the
pricing model as well as the optimal redeeming strategy. In particular, if
the accumulative dividends are assumed to be returned to the borrower on
redemption, the pricing model will get one path-dependent variable involved,
which leads the study of the associated redeeming strategy to be
challenging. We will provide an analytical method to analyze the optimal
redeeming strategy.

Throughout the paper, we assume that the risk neutral stock price follows a
geometric Brownian motion:%
\[
dS_{t}=\left( r-\delta \right) S_{t}dt+\sigma S_{t}dW_{t},
\]%
where constants $r>0,$ $\delta \geq 0$ and $\sigma >0$ are the riskless
interest rate, continuous dividend yield\footnote{%
All models can be trivially extended to the discrete dividend payment. On
the other hand, the assumption of continuous dividend payments is true if
the collateralized security is foreign currency.} of and volatility of the
stock, respectively, and $\{W_{t};t>0\}$ is a standard 1-dimension Brownian
motion on a filtered probability space $(\mathbb{S},\mathscr F,\{\mathscr %
F_{t}\}_{t\geq 0},\mathbb{P})$ with $W_{0}=0$ almost surely. We denote by $%
K>0$ the stock loan's principal, $\gamma $ the (continuous compounding) loan
interest rate, and $T>0$ the maturity date. For later use, we let $%
C_{E}(\cdot ,\cdot ;r,\delta ,X)$ (or $P_{E}(\cdot ,\cdot ;r,\delta ,X)$) be
the price of European vanilla call (or put) option with riskless rate $r,$
dividend yield $\delta $ and strike price $X.$

The rest of the paper is organized as follows. In section 2, we consider a
relatively simple case: the dividends are gained by the lender before
redemption. In section 3, we assume that the dividends are reinvested in the
stock and will be returned to the borrower on redemption. We will see that
this reduces to a special case in section 2, and the resulting optimal
strategy can be linked to those presented in the subsequent two sections.
Section 4 is devoted to the scenario that the cash dividends are delivered
to the borrower immediately, no matter whether the borrower redeems the
stock. It is worth pointing out that the pricing models in Section 2-4
involve one state variable only and are then relatively easy to analyze. In
section 5, we will take into consideration the most challenging and
interesting case: the accumulative cash dividends are delivered to the
borrower on redemption. The pricing model proves to be a two-dimensional
parabolic variational inequality. Section 6 discusses some extensions with
conclusive remarks.

\section{Dividends gained by the lender before redemption}

\setcounter{equation}{0} \setcounter{theorem}{0}Assume that the dividends
are gained by the lender before redemption. Let $V_{1}=V_{1}(S,t)$ be the
value of the stock loan at time $t$ with stock price $S$. Then
\[
V_{1}(S_{t},t)=\sup_{v\in \mathcal{T}_{[t,T]}}\mathbb{E}_{t}\left[
e^{-r(v-t)}(S_{v}-Ke^{\gamma v})^{+}\right] .
\]%
where $\mathbb{E}_{t}$ is the risk neutral expectation conditionally on $%
\mathscr F_{t}$ and $\mathcal{T}_{[t,T]}$ denotes the set of all $\left\{
\mathcal{F}_{s}\right\} _{t\leqslant s\leqslant T}$-stopping times with
values in $[t,T].$ It turns out that $V_{1}$ satisfies the following
variational inequality [cf. Karatzas and Shreve (1998)]:
\begin{equation}
\left\{
\begin{array}{l}
\min \left\{ -L_{S}^{r,\delta }V_{1},V_{1}-\left( S-Ke^{\gamma t}\right)
\right\} =0, \\
V_{1}(S,T)=\left( S-Ke^{\gamma T}\right) ^{+},\hspace{1in}(S,t)\in Q,%
\end{array}%
\right.  \label{eq0}
\end{equation}%
where $Q\equiv \left( 0,\infty \right) \times \lbrack 0,T),$ and
\[
L_{S}^{r,\delta }=\frac{\partial }{\partial t}+\frac{1}{2}\sigma ^{2}S^{2}%
\frac{\partial ^{2}}{\partial S^{2}}+\left( r-\delta \right)S
\frac{\partial }{\partial S}-r.
\]

Let us define the redemption region
\[
E_1\equiv \{\left( S,t\right) \in Q:V_1(S,t)=S-Ke^{\gamma t}\}.
\]
The following proposition characterizes the properties of $E_1.$

\begin{proposition}
\label{trivial}Assume that the dividends are gained by the lender
before redemption.

i) If $r\geq \gamma $ and $\delta =0,$ then $E_1=\emptyset .$ This
indicates that early redemption should never happen. In this
scenario the stock loan is equivalent to a European call option
with strike price $Ke^{\gamma T}$, namely,
$V_1(S,t)=C_E(S,t;r,0,Ke^{\gamma T}).$

ii) If $r\geq \gamma $ and $\delta >0,$ or $r<\gamma ,$ then there
is an optimal redeeming boundary $S_1^{*}(t):[0,T)\rightarrow
(0,\infty )$ such that
\[
E_1=\{\left( S,t\right) \in Q:S\geq S_1^{*}(t)\}.
\]
In addition, $e^{-\gamma t}S_1^{*}(t)$ is monotonically decreasing with $t,$
\begin{equation}
S_1^{*}(T)\equiv \lim_{t\rightarrow T{-}}S_1^{*}(t)=\left\{
\begin{array}{l}
e^{\gamma T}\max \left( K,\frac{r-\gamma }\delta K\right) \text{ if }r\geq
\gamma \text{ and }\delta >0,\text{ } \\
e^{\gamma T}K\text{ if }r<\gamma .
\end{array}
\right.   \label{eqnew}
\end{equation}
\end{proposition}

\noindent Proof: By similarity reduction
\[
f_1(x,t)=e^{-\gamma t}V_1(S,t),\text{ }x=e^{-\gamma t}S,
\]
we get
\begin{equation}
\left\{
\begin{array}{l}
\min \left\{ -L_x^{\overline{r},\delta }f_1,f_1-\left( x-K\right) \right\}
=0, \\
f_1(x,T)=\left( x-K\right) ^{+},\hspace{1.0in}\text{ }(x,t)\in Q,%
\end{array}
\right.  \label{eq1}
\end{equation}
where $\overline{r}=r-\gamma $. Then, $f_1(x,t)$ can be regarded as the
value of an American call option with riskless rate $\overline{r}$ that is
likely to be non-positive. Hence, part i) is a well-known result when $%
\overline{r}>0$ and $\delta =0$.

When $\overline{r}\geq 0$ and $\delta >0,$ we know from (\ref{eq1}) [cf.
Jiang (2005)] that there is a decreasing function $x_{1}^{\ast
}(t):[0,T)\rightarrow (0,+\infty ),$ such that
\begin{equation}
\left\{ \left( x,t\right) \in Q:f_{1}(x,t)=x-K\right\} =\left\{ \left(
x,t\right) \in Q:x\geq x_{1}^{\ast }(t),\text{ }t\in \lbrack 0,T)\right\}
\label{eq11}
\end{equation}%
and
\begin{equation}
x_{1}^{\ast }(T)\equiv \lim_{t\rightarrow T-}x_{1}^{\ast }(t)=\max \left( K,%
\frac{\overline{r}}{\delta }K\right) .  \label{eq12}
\end{equation}%
Using a similar argument as in Jiang (2005), we are able to show that (\ref%
{eq11}) and (\ref{eq12}) are still true when $\overline{r}<0.$ Notice that $%
\max \left( K,\frac{\overline{r}}{\delta }K\right) =K$ for $\overline{r}<0,$
part ii) then follows.\qed

\hspace{1.0in}

Now we examine the asymptotic behavior of $S_1^{*}(t)$ as the time to
maturity goes to infinity.

\begin{proposition}
\label{xz}Assume $r\geq \gamma $ and $\delta >0,$ or $r<\gamma .$ Let $%
S_1^{*}(t)$ be as given in part ii) of Proposition \ref{trivial}. Then
\begin{equation}
S_{1,\infty }^{*}\equiv \lim_{T\rightarrow +\infty }S_1^{*}(t)=e^{\gamma
t}x_{1,\infty }^{*},  \label{eqinf}
\end{equation}
where
\begin{equation}
x_{1,\infty }^{*}=\left\{
\begin{array}{l}
\frac{\alpha _{+}}{\alpha _{+}-1}K,\text{ if }\delta >0,\text{ or }\delta =0%
\text{ and }r<\gamma -\frac 12\sigma ^2 \\
+\infty ,\text{\ \ if }\delta =0\text{ and }\gamma -\frac 12\sigma
^2\leq r<\gamma
\end{array}
\right.   \label{px}
\end{equation}
and
\begin{equation}
\alpha _{+}=\frac{-\left( r-\gamma -\delta -\frac 12\sigma ^2\right) +\sqrt{%
\left( r-\gamma -\delta +\frac 12\sigma ^2\right) ^2+2\delta \sigma ^2}}{%
\sigma ^2}.  \label{alpha}
\end{equation}
\end{proposition}

To prove the above proposition, we only need to consider a perpetual stock
loan and $S_{1,\infty }^{*}$ is the corresponding optimal redeeming
boundary. This has been done by Xia and Zhou (2007) in terms of a
probabilistic approach. We would like to provide a simple PDE argument which
is placed in Appendix. It is worth pointing out that the explicit solution
of $x_{1,\infty }^{*}$ is nothing but the optimal exercise boundary of a
perpetual American call option when $r\geq \gamma $ and $\delta >0.$

\section{Reinvested dividends returned to the borrower on redemption}

\setcounter{equation}{0} \setcounter{theorem}{0}

Assume that the dividends are immediately re-invested in the stock and will
be returned to the borrower on redemption. The intrinsic value (i.e. the
redemption payoff) of the stock loan at time $t$ becomes
\[
\left( e^{\delta t}S_{t}-Ke^{\gamma t}\right) ^{+},\text{ }t\in \lbrack
0,T].
\]%
Let $V_{2}=V_{2}(S,t)$ be the price function of the stock loan. Then
\begin{equation}
V_{2}(S_{t},t)=\sup_{v\in \mathcal{T}_{[t,T]}}\mathbb{E}_{t}\left[
e^{-r(v-t)}(e^{\delta v}S_{v}-Ke^{\gamma v})^{+}\right] .  \label{sec2}
\end{equation}

Denote $\widehat{V}_{2}(\widehat{S}_{t},t)=V_{2}(S_{t},t),$ where
\[
\widehat{S}_{t}\equiv e^{\delta t}S_{t}=S_{0}\exp \left\{ \left( r-\frac{%
\sigma ^{2}}{2}\right) t+\sigma W_{t}\right\}.
\]%
Then we can rewrite (\ref{sec2}) as
\begin{equation}
\widehat{V}_{2}(\widehat{S}_{t},t)=\sup_{v\in \mathcal{T}_{[t,T]}}\mathbb{E}%
_{t}\left[ e^{-r(v-t)}(\widehat{S}_{v}-Ke^{\gamma v})^{+}\right] ,
\label{v0}
\end{equation}%
satisfying
\begin{equation}
\left\{
\begin{array}{l}
\min \left\{ -L_{\widehat{S}}^{r,0}\widehat{V}_{2},\widehat{V}_{2}-\left(
\widehat{S}-Ke^{\gamma t}\right) \right\} =0, \\
\widehat{V}_{2}(S,T)=\left( \widehat{S}-Ke^{\gamma T}\right) ^{+},\hspace{1in%
}(\widehat{S},t)\in Q.%
\end{array}%
\right.  \label{v01}
\end{equation}%
As a result, the stock loan can be regarded as the one written on the
non-dividend paying stock $\widehat{S}$, which has been studied in Section 2.

For later use, we define the redemption region associated with $\widehat{V}%
_2 $ as
\begin{equation}
\widehat{E}_2=\left\{ \left( \widehat{S},t\right) \in Q:\widehat{V}_2(%
\widehat{S},t)=\widehat{S}-Ke^{\gamma t}\right\} .  \label{v1}
\end{equation}
Later we will see that $\widehat{E}_2$ has a close link with the redemption
regions addressed in subsequence two sections. By Proposition \ref{trivial},
$\widehat{E}_2=\emptyset $ when $r\geq \gamma ,$ and if $r<\gamma ,$ then
there is a monotonically decreasing function $\widehat{S}_2^{*}(t):[0,T)%
\rightarrow (0,\infty )$ such that
\begin{equation}
\widehat{E}_2=\left\{ \left( \widehat{S},t\right) \in Q:\widehat{S}\geq
\widehat{S}_2^{*}(t),\text{ }t\in [0,T)\right\} ,  \label{vs}
\end{equation}
$\widehat{S}_2^{*}(T)\equiv \lim_{t\rightarrow T{-}}\widehat{S}_2^{*}(t)=K,$
and
\begin{equation}
\widehat{S}_{2,\infty }^{*}\equiv \lim_{T\rightarrow {+\infty }}\widehat{S}%
_2^{*}(t)\text{ is finite if }r<\gamma -\frac 12\sigma ^2,\text{and infinite
if }\gamma -\frac 12\sigma ^2\leq r<\gamma .  \label{end0}
\end{equation}

\section{Dividends always delivered to the borrower}

\setcounter{equation}{0} \setcounter{theorem}{0}Assume that the dividends
are always delivered to the borrower during the lifetime of the stock loan.
The intrinsic value of the stock loan becomes
\begin{equation}
\left( S_{t}-Ke^{\gamma t}\right) ^{+}+\int_{0}^{t}\delta e^{r(t-u)}S_{u}du,%
\text{ }t\in \lbrack 0,T].  \label{intr}
\end{equation}%
By introducing a path-dependent variable
\begin{equation}
I_{t}=\int_{0}^{t}\delta e^{r(t-u)}S_{u}du,  \label{path}
\end{equation}%
the value of the stock loan can be expressed as
\begin{eqnarray*}
V_{3}(S_{t},I_{t},t) &=&\sup_{v\in \mathcal{T}_{[t,T]}}\mathbb{E}_{t}\left(
e^{-r(v-t)}\left[ (S_{v}-Ke^{\gamma v})^{+}+I_{v}\right] \right) \\
&=&I_{t}+\sup_{v\in \mathcal{T}_{[t,T]}}\mathbb{E}_{t}\left( e^{-r(v-t)}%
\left[ (S_{v}-Ke^{\gamma v})^{+}\right] +\int_{t}^{v}\delta
e^{-r(u-t)}S_{u}du\right) .
\end{eqnarray*}

Observe that $V_{3}(S_{t},I_{t},t)-I_{t}$ is independent of $I_{t}.$ Then we
can write $H(S,t)\equiv V_{3}(S,I,t)-I,$ which satisfies
\begin{equation}
\left\{
\begin{array}{l}
\min \left\{ -L_{S}^{r,\delta }H-\delta S,H-\left( S-Ke^{\gamma t}\right)
\right\} =0,\text{ } \\
H(S,T)=\left( S-Ke^{\gamma T}\right) ^{+},\hspace{1in}(S,t)\in Q.%
\end{array}%
\right.  \label{mod2}
\end{equation}%
Here $H(S,t)$ can be regarded as the value of the stock loan excluding the
accumulative dividends. Compared with (\ref{eq0}) in Section 2, (\ref{mod2})
has a source term $\delta S$ due to the dividends delivered.

Now we define the redemption region
\[
E_{3}\equiv \{(S,t)\in Q:H(S,t)=S-Ke^{\gamma t}\}.
\]%
We proceed with a lemma.

\begin{lemma}
\label{new}Let $\widehat{V}_2(\cdot ,\cdot )$ and $\widehat{E}_2$ be as
given in (\ref{v0}) and (\ref{v1}) respectively. Then
\[
\widehat{V}_2(S,t)\leq H(S,t)\text{ and }E_3\subset \widehat{E}_2.
\]
\end{lemma}

\noindent Proof: First we prove $\widehat{V}_2(S,t)\leq H(S,t).$ Compared
with (\ref{v01}), (\ref{mod2}) can be rewritten as
\[
\left\{
\begin{array}{l}
\min \left\{ -L_S^{r,0}H-\delta S\left( 1-\frac{\partial H}{\partial S}%
\right) ,H-\left( S-Ke^{\gamma t}\right) \right\} =0,\text{ } \\
H(S,T)=\left( S-Ke^{\gamma T}\right) ^{+},\hspace{1.0in}(S,t)\in Q.%
\end{array}
\right.
\]
So, by the maximum principle [cf. Friedman (1988)], it suffices to show $%
\frac{\partial H}{\partial S}\leq 1$ or
\begin{equation}
\frac \partial {\partial S}(H-S)\leq 0.  \label{end}
\end{equation}
Notice
\[
\left\{
\begin{array}{l}
\min \left\{ -L_S^{r,\delta }\left( H-S\right) ,\left( H-S\right)
+Ke^{\gamma t}\right\} =0,\text{ } \\
H(S,T)-S=\max (-S,-Ke^{\gamma T}),\hspace{1.0in}(S,t)\in Q.%
\end{array}
\right.
\]
Again applying the maximum principle gives (\ref{end}), then $\widehat{V}%
_2(S,t)\leq H(S,t)$ follows.

Now let us show $E_3\subset \widehat{E}_2.$ For any $(S,t)\in E_3,$ we have
\[
S-Ke^{\gamma t}\leq \widehat{V}_2(S,t)\leq H(S,t)=S-Ke^{\gamma t},
\]
which implies $\widehat{V}_2(S,t)=S-Ke^{\gamma t},$ i.e., $(S,t)\in \widehat{%
E}_2.$ This completes the proof.\qed

\hspace{1.0in}

The following proposition characterizes the shape of $E_3.$

\begin{proposition}
\label{del}Assume $\delta >0$ and the dividends are always delivered to the
borrower during the lifetime of the stock loan.

i) If $r\geq \gamma ,$ then $E_3=\emptyset .$ That is, early
redemption should never happen. In addition, we have
\begin{equation}
H(S,t)=C_E(S,t;r,\delta ,Ke^{\gamma T})+(1-e^{-\delta (T-t)})S,
\label{parity}
\end{equation}

ii) If $r<\gamma ,$ then there is an optimal redeeming boundary $%
S_3^{*}(t):[0,T)\rightarrow (0,+\infty )$ such that
\[
E_3=\{\left( S,t\right) \in Q:S\geq S_3^{*}(t)\}.
\]
In addition, $e^{-\gamma t}S_3^{*}(t)$ is monotonically decreasing in $t,$
\[
S_3^{*}(T)\equiv \lim_{t\rightarrow T{-}}S_3^{*}(t)=e^{\gamma T}K,
\]
and
\begin{equation}
S_3^{*}(t)\geq \widehat{S}_2^{*}(t),  \label{end2}
\end{equation}
where $\widehat{S}_2^{*}(t)$ is as given in (\ref{vs}).
\end{proposition}

\noindent Proof: First, by an arbitrage argument, it is not hard to get
\begin{equation}
H(S,t)\geq C_E(S,t;r,\delta ,Ke^{\gamma T})+(1-e^{-\delta (T-t)})S,
\label{imp0}
\end{equation}
where the right hand side is the price of the corresponding stock loan
without early redemption right. Since $C_E(S,t;r,\delta ,Ke^{\gamma
T})>Se^{-\delta (T-t)}-Ke^{\gamma T}e^{-r(T-t)}$ [see, for example, Hull
(2003)], it follows
\begin{eqnarray*}
H(S,t) &\geq &S-Ke^{\gamma T}e^{-r(T-t)} \\
&>&S-Ke^{\gamma t},\text{ for }r\geq \gamma \text{ and }t<T,
\end{eqnarray*}
which implies part i).

To show part ii), as before, we make a transformation: $f_3(x,t)=e^{-\gamma
t}H(S,t),\text{ }x=e^{-\gamma t}S. $ It follows
\begin{equation}
\left\{
\begin{array}{l}
\min \left\{ -L_x^{\overline{r},\delta }f_3-\delta x,f_3-(x-K)\right\} =0,%
\text{ } \\
f_3(x,T)=\left( x-K\right) ^{+},\hspace{1.0in}(x,t)\in Q.%
\end{array}
\right.  \label{daim}
\end{equation}
By (\ref{end}), we have $\frac \partial {\partial x}(f_3-x)\leq 0,$
from
which we infer that there is a single-value function $x_3^{*}(t):[0,T)%
\rightarrow $ $(0,+\infty )\cup +\infty $ such that
\[
\left\{ \left( x,t\right) \in Q:f_3(x,t)=x-K\right\} =\{\left( x,t\right)
\in Q:x\geq x_3^{*}(t)\}.
\]
Since
\[
-L_x^{\overline{r},\delta }\left( x-K\right) -\delta x=-\overline{r}K>0,%
\text{ for }r<\gamma ,
\]
using a similar analysis as in Brezis and Friedman (1976) or Dai et al.
(2004), we can deduce $f_3(x,t)-(x-K)$ has a compact support for all $t.$
This indicates $x^{*}(t)<+\infty$ for all $t$. So, we can take $%
S_3^{*}(t)=e^{\gamma t}x_3^{*}(t).$ (\ref{end2}) is a corollary of Lemma \ref%
{new}.

Apparently
\begin{equation}
\frac{\partial f_{3}}{\partial t}\leq 0,  \label{imp}
\end{equation}%
which yields the monotonicity of $x_{3}^{\ast }(t).$ It remains to show $%
x_{3}^{\ast }(T)\equiv \lim_{t\rightarrow T{-}}x_{3}^{\ast }(t)=K.$ The
argument is standard and is stated as follows. Obviously $x_{3}^{\ast
}(t)\geq K.$ If $x_{3}^{\ast }(T)>K,$ we would have for $\widetilde{x}\in
\left( K,x_{3}^{\ast }(T)\right) ,$%
\[
\left. \frac{\partial f_{3}}{\partial t}\right\vert _{(\widetilde{x}%
,T)}=\left. \left[ -\frac{1}{2}\sigma ^{2}x^{2}\frac{\partial ^{2}}{\partial
x^{2}}-\left( \overline{r}-\delta \right) x\frac{\partial }{\partial x}+%
\overline{r}\right] \left( x-K\right) -\delta x\right\vert _{x=\widetilde{x}%
}=-\overline{r}K>0,
\]%
which is in contradiction with (\ref{imp}). This completes the proof.\qed

\hspace{1.0in}

Now we analyze the asymptotic behavior of $S_3^{*}(t)$ as the time to
maturity goes to infinity. By virtue of (\ref{end0}) and (\ref{end2}), we
immediately obtain that $S_3^{*}(t)\rightarrow +\infty $ as $T\rightarrow
+\infty $ in the case of $\gamma -\frac 12\sigma ^2\leq r<\gamma .$ We will
show that it is also true for $r<\gamma -\frac{\sigma ^2}2.$ Again, we need
to study the corresponding perpetual stock loan. Denote $H_\infty
(S)=\lim_{T\rightarrow +\infty }H(S,t).$ We assert
\begin{equation}
H_\infty (S)=S.  \label{imp10}
\end{equation}
Indeed, from (\ref{imp0}), we have
\[
H(S,t)\geq S-Se^{-\delta (T-t)},
\]
which yields $H_\infty (S)\geq S$ by letting $T\rightarrow \infty.$
The converse inequality is apparent. So, we get (\ref{imp10}), which
implies that one should never redeem the perpetual stock loan. We
summarize the above result as follows:

\begin{proposition}
Assume $\delta >0$ and the dividends are always delivered to the borrower
during the lifetime of the stock loan. Then the perpetual stock loan is
equivalent to the stock, i.e., $H_\infty (S)\equiv \lim_{T\rightarrow
+\infty }H(S,t)=S$. In addition,
\[
S_{3,\infty }^{*}\equiv \lim_{T\rightarrow +\infty }S_3^{*}(t)=+\infty ,%
\text{ for }r<\gamma ,
\]
where $S_3^{*}(t)$ is the optimal redeeming boundary in part ii)
of Proposition \ref{del}.
\end{proposition}

\section{Dividends returned to the borrower on redemption}

\setcounter{equation}{0} \setcounter{theorem}{0} In this section, we assume
that the accumulative dividends will be returned to the borrower on
redemption. In contrast to (\ref{intr}), the intrinsic value of the stock
loan is now
\[
\left( S_{t}-Ke^{\gamma t}+\int_{0}^{t}\delta e^{r(t-u)}S_{u}du\right) ^{+},%
\text{ }t\in \lbrack 0,T].
\]

Again we introduce the path-dependent variable $I_{t}$ as given in (\ref%
{path}), then the value of the stock loan can be expressed as
\[
V_{4}(S_{t},I_{t},t)=\sup_{v\in \mathcal{T}_{[t,T]}}\mathbb{E}_{t}\left[
e^{-r(v-t)}(S_{v}-Ke^{\gamma v}+I_{v})^{+}\right] .
\]%
Note that
\[
dI_{t}=\left( \delta S_{t}+rI_{t}\right) dt.
\]%
It follows that $V_{4}(S,I,t)$ satisfies
\begin{equation}
\left\{
\begin{array}{l}
\min \left\{ -L_{S}^{r,\delta }V_{4}-\left( \delta S+rI\right) \frac{%
\partial V_{4}}{\partial I},V_{4}-\left( S+I-Ke^{\gamma t}\right) \right\}
=0, \\
V_{4}(S,I,T)=\left( S+I-Ke^{\gamma T}\right) ^{+},\text{ }\hspace{1in}%
(S,I,t)\in \Omega ,%
\end{array}%
\right. \text{ }  \label{model}
\end{equation}%
where $\Omega =(0,+\infty )\times (0,+\infty )\times \lbrack 0,T).$ (\ref%
{model}) is analogous to the pricing model for Asian options and the
existence of (strong) solution can be proved using a similar argument as in
Pascucci (2008). We emphasize that (\ref{model}) is indeed a two-dimensional
time-dependent problem and does not permit dimension reduction.

As before, we define the redemption region by
\[
E_{4}=\left\{ \left( S,I,t\right) \in \Omega :V_{4}(S,I,t)=S+I-Ke^{\gamma
t}\right\} .
\]%
Apparently $V_{4}(S,I,t)\leq V_{3}(S,I,t)=H(S,t)+I,$ from which we
immediately get
\[
\left\{ \left( S,I,t\right) \in \Omega :(S,t)\in E_{3}\right\} \subset
E_{4}.
\]%
The following lemma presents a stronger result which plays an important role
in analysis of the shape of $E_{4}$:

\begin{lemma}
\label{com2}Let $\widehat{V}_2(\cdot ,\cdot )$ and $\widehat{E}_2$ be as
defined in (\ref{v0}) and (\ref{v1}) respectively. Then we have
\begin{equation}
V_4(S,I,t)\leq \widehat{V}_2(S+I,t)  \label{inter}
\end{equation}
and thus
\begin{equation}
\left\{ \left( S,I,t\right) \in \Omega :(S+I,t)\in \widehat{E}_2\right\}
\subset E_4.  \label{comp2}
\end{equation}
\end{lemma}

\noindent Proof: We only need to prove $V_{4}(S,I,t)\leq \widehat{V}%
_{2}(S+I,t).$ Let us adopt $y\equiv S+I$ as a new state variable in place of
$S,$ and denote

\[
U(y,I,t)=V_4(S,I,t),
\]
which satisfies
\[
\left\{
\begin{array}{l}
\min \left\{ -\mathcal{L}U,U-\left( y-Ke^{\gamma t}\right) \right\} =0, \\
U(y,I,T)=\left( y-Ke^{\gamma T}\right) ^{+},\hspace{1.0in}(y,I,t)\in \Omega
_y,%
\end{array}
\right. \text{ }
\]
where $\Omega _y\equiv \{(y,I,t):\ 0<I<y<+\infty,\ 0\leq t<T\},$
\[
\mathcal{L}=\frac \partial {\partial t}+\left( \delta y+\left( r-\delta
\right) I\right) \frac \partial {\partial I}+\frac 12\sigma ^2\left(
y-I\right) ^2\frac{\partial ^2}{\partial y^2}+ry\frac \partial {\partial
y}-r.
\]

Note that (\ref{v01}) can be rewritten as
\[
\left\{
\begin{array}{l}
\min \left\{ -\mathcal{L}\widehat{V}_2-\frac 12\sigma ^2\left[ y^2-(y-I)^2%
\right] \frac{\partial ^2\widehat{V}_2}{\partial y^2},\widehat{V}_2-\left(
y-Ke^{\gamma t}\right) \right\} =0, \\
\widehat{V}_2(y,T)=\left( y-Ke^{\gamma T}\right) ^{+},\hspace{1.0in}%
(y,I,t)\in \Omega _y.%
\end{array}
\right. \text{ }
\]
Due to the convexity of $\left( y-Ke^{\gamma t}\right) ^{+}$ in $y,$ we can
deduce $\frac{\partial ^2\widehat{V}_2}{\partial y^2}\geq 0,$ then
\[
\frac 12\sigma ^2\left[ y^2-(y-I)^2\right] \frac{\partial ^2\widehat{V}_2}{%
\partial y^2}\geq 0,\text{ for any }y>I>0,\text{ }t\in [0,T).
\]
Applying the maximum principle then gives
\[
U(y,I,t)\leq \widehat{V}_2(y,t),
\]
which is desired.\qed

\hspace{1.0in}

We would like to give a financial interpretation to (\ref{inter}). $\widehat{%
V}_{2}(\cdot ,\cdot )$ and $V_{4}(\cdot ,\cdot ,\cdot )$ represent the
prices for the stock loans respectively with reinvested dividends and with
cash dividends, both of which will be returned to the borrowers on
redemption. In the risk-neutral world, the return rate of reinvested
dividends is the same as that of cash dividends. Since we can regard the
combination of the dividends and the stock as an imaginary underlying asset
(i.e. $y),$ the cash dividends essentially decrease the volatility of the
underlying asset, which leads to a lower price of the corresponding stock
loan.

\subsection{The case of $r\geq \protect\gamma $}

Now let us investigate the case of $r\geq \gamma $.
\begin{proposition}
\label{pp}Assume $\delta >0$ and the dividends are gained by the
borrower on redemption.

i) If $r>\gamma ,$ then $E_3=\emptyset .$ That is, early
redemption should never happen.

ii) If $r=\gamma ,$ then it is optimal to hold the option before expiry.
\end{proposition}

\noindent Proof: We adopt an arbitrage argument. If the borrower redeems the
loan at time $t<T,$ then he or she will get
\[
\text{one stock }S_t+\text{ cash }I_t-Ke^{\gamma t},
\]
which amounts to at expiry
\begin{eqnarray}
\ \ S_T+I_T-Ke^{\gamma t}e^{r(T-t)} &<&S_T+I_T-Ke^{\gamma T}  \label{ine} \\
&\leq &\left( S_T+I_T-Ke^{\gamma T}\right) ^{+},  \nonumber
\end{eqnarray}
This indicates that early redemption is not optimal. Part i) then follows.

To prove part ii), the argument is similar and the unique difference lies in
that ``$<$'' in (\ref{ine}) should be replaced by ``$=".$\qed

\hspace{1.0in}

It is worth distinguishing two statements of part i) and ii) in Proposition %
\ref{pp}. For the latter case, early redemption may be optimal on some
occasions. For example, when $I\geq Ke^{rt}$ and $r=\gamma ,$ it is not hard
to show that $V_{4}(S,I,t)=S+I-Ke^{rt},$ which implies that the redemption
at any time is optimal.

\hspace{1.0in}

Owing to Proposition \ref{pp}, (\ref{model}) is reduced to a linear problem:
\[
\left\{
\begin{array}{l}
-L_S^{r,\delta }V_4-\left( \delta S+rI\right) \frac{\partial V_4}{\partial I}%
=0, \\
V_4(S,I,T)=\left( S+I-Ke^{\gamma T}\right) ^{+},\hspace{.5in}(S,I,t)\in
\Omega .%
\end{array}
\right. \text{ }
\]

\subsection{The case of $r<\protect\gamma $}

By Lemma \ref{com2} and the properties of $\widehat{E}_2$, we know that $E_4$
is always non-empty when $r<\gamma .$ Furthermore, we have

\begin{proposition}
\label{pn}Assume $\delta >0$ and the dividends are gained by the
borrower on redemption. If $r<\gamma ,$ then $\{\left(
S,I,t\right) \in \Omega :I\geq Ke^{\gamma t}\}\subset E_4$.
\end{proposition}

\noindent Proof: Given $\left( S_{t},I_{t},t\right) \in \Omega $ with $%
I_{t}\geq Ke^{\gamma t},$ we claim that the loan should be redeemed
immediately at time $t.$ Indeed, if it is redeemed at time $t,$ then we have
a payoff: stock $S_{t}+$ non-negative cash $I_{t}-Ke^{\gamma t},$ which
becomes at a later time $t^{\prime }\in (t,T]$%
\[
\text{stock }S_{t^{\prime }}+\text{cash }I_{t^{\prime }}-Ke^{\gamma
t}e^{r(t^{\prime }-t)}>\left( S_{t^{\prime }}+I_{t^{\prime }}-Ke^{\gamma
t^{\prime }}\right) ^{+}\text{ if }r<\gamma .
\]%
This implies the conclusion.\qed

\hspace{1.0in}

We stress that Proposition \ref{pp} and Lemma \ref{pn} only rely on the
no-arbitrage principle, and therefore are independent of the geometric
Brownian motion assumption of stock price. Such a remark also applies to
some of previous results.

\hspace{1.0in}

Denote $\widehat{\Omega }=\{\left( S,I,t\right) \in \Omega :I<Ke^{\gamma
t}\}.$ Due to (\ref{pn}), we only need to study
\[
\widehat{E}_4\equiv E_4\cap \widehat{\Omega }.
\]

\begin{proposition}
\label{main1}Assume that the dividends are gained by the borrower on
redemption and $r<\gamma .$ Then, there is an optimal redeeming boundary $%
S_4^{*}(I,t):(0,Ke^{\gamma t})\times [0,T)$ $\rightarrow (0,\infty )$ such
that
\[
\widehat{E}_4=\{\left( S,I,t\right) \in \widehat{\Omega }:S\geq
S_4^{*}(I,t)\}.
\]
Moreover, $e^{-\gamma t}S_4^{*}(I,t)$ is monotonically decreasing in $I$ and
$t$,
\[
S_4^{*}(I,T)\equiv \lim_{t\rightarrow T}S_4^{*}(I,t)=e^{\gamma T}K-I,\text{
for all }I,
\]
and
\begin{equation}
S_4^{*}(I,t)\leq \widehat{S}_2^{*}(t)-I, \text{ for all }t,
\label{end3}
\end{equation}
where $\widehat{S}_2^{*}(t)$ is as given in (\ref{vs}). In
particular, $S_4^{*}(0,t)\equiv \lim_{I\rightarrow
0}S_4^{*}(I,t)\leq \widehat{S}_2^{*}(t),\text{ for all }t$.
\end{proposition}

\noindent Proof: Since $(S+I-Ke^{\gamma t})^{+}$ is convex in $S$, we infer
that $V_4(S,I,t)$ is also convex in $S,$ namely
\[
V_4(aS_1+\left( 1-a\right) S_2,I,t)\leq aV_4(S_1,I,t)+\left( 1-a\right)
V_4\left( S_2,I,t\right) .
\]
which implies the convexity of $\widehat{E}_4$ in $S$. On other hand, due to
Lemma \ref{com2}, we have $(S,I,t)\in \widehat{E}_4$ for $S>\widehat{S}%
_2^{*}(t)-I$. This indicates the existence of $S_4^{*}(I,t)$ and $%
S_4^{*}(I,t)\leq \widehat{S}_2^{*}(t)-I.$

Using the similarity reduction $V_4(S,I,t)=e^{-\gamma t}f_4(x,A,t),$ $%
x=e^{-\gamma t}S$ and $A=e^{-\gamma t}I,$ we get
\begin{equation}
\left\{
\begin{array}{l}
\min \left\{ -L_x^{\overline{r},\delta }f_4-\left( \delta x+\overline{r}%
A\right) \frac{\partial f_4}{\partial A},f_4-\left( x+A-K\right) \right\} =0,%
\text{ } \\
f_4(x,A,T)=\left( x+A-K\right) ^{+},\hspace{1.0in}\left( x,A,t\right) \in
\Omega .%
\end{array}
\right.  \label{gpro}
\end{equation}
Let $x_4^{*}(A,t)\equiv e^{-\gamma t}S_4^{*}(I,t).$ It suffices to show that
$x_4^{*}(A,t)$ is monotonically decreasing in $A$ and $t,$ and
\begin{equation}
x_4^{*}(A,T)\equiv \lim_{t\rightarrow T}x_4^{*}(A,t)=K-A,\text{ for all }A.
\label{gp1}
\end{equation}
By the maximum principle, it is not hard to get
\[
\frac{\partial f_4}{\partial t}\leq 0,\text{ }
\]
and
\[
\frac \partial {\partial A}\left[ f_4-(x+A-K)\right] =\frac{\partial f_4}{%
\partial A}-1\leq 0,
\]
which implies the monotonicity of $x_4^{*}(A,t)$ in $A$ and $t.$ Using a
similar argument as in Proposition \ref{del}, we are able to obtain (\ref%
{gp1}).\qed

\section{Numerical examples}

In this section we present numerical results to verify our theoretical
results. Let us first look at the pricing models in Section 2-4 which belong
to standard one-dimensional parabolic variational inequalities. They can be
numerically solved using many sophisticated numerical methods such as the
projected SOR [cf. Wilmott et al. (1993)], the recursive integration method
[cf. Huang et al. (1996)] and the penalty approach [cf. Forsyth and Vetzal
(2002)]. We simply make use of the binomial tree method [cf. Hull (2003)]
which is easy to implement. The default data used are $r-\gamma =-0.04,$ $%
\delta =0.03,$ $\gamma =0.1$ and $K=0.7.$ Figure 1 and Figure 2 plot the
optimal redeeming boundaries against the time to maturity $\tau =T-t$ with $%
\sigma =0.4$ and with $\sigma =0.15,$ respectively. Here $x_1^{*}\left(
\cdot \right) =e^{-\gamma t}S_1^{*}(t),$ $x_2^{*}\left( \cdot \right)
=e^{-\gamma t}\widehat{S}_2^{*}(t)$ and $x_3^{*}\left( \cdot \right)
=e^{-\gamma t}S_3^{*}(t).$ Observe that $x_1^{*}\leq x_2^{*}\leq x_3^{*},$
and these boundaries are monotonically increasing in $\tau $ (so, decreasing
in $t$). Since $r$ $<\gamma ,$ we can see that all boundaries go to $%
x=K\equiv 0.7$ at $\tau \rightarrow 0$. Theoretical results indicate that $%
x_1^{*}\left( \cdot \right) $ has an asymptotic line $(x=1.97$ for Figure 1
and $x=0.82$ for Figure 2) while $x_3^{*}\left( \cdot \right) $ does not as $%
\tau $ $\rightarrow +\infty ,$ which can be seen from the figures. In
addition, $x_2^{*}\left( \cdot \right) $ has an asymptotic line $(x=0.97)$
as $\tau \rightarrow +\infty $ in Figure 2 due to $r<\gamma -\frac{\sigma ^2}%
2$, whereas not in Figure 1 ($\gamma -\frac{\sigma ^2}2\leq r<\gamma $).
This is also consistent with our theoretical analysis.

\[
\begin{array}{c}
\psfig{figure=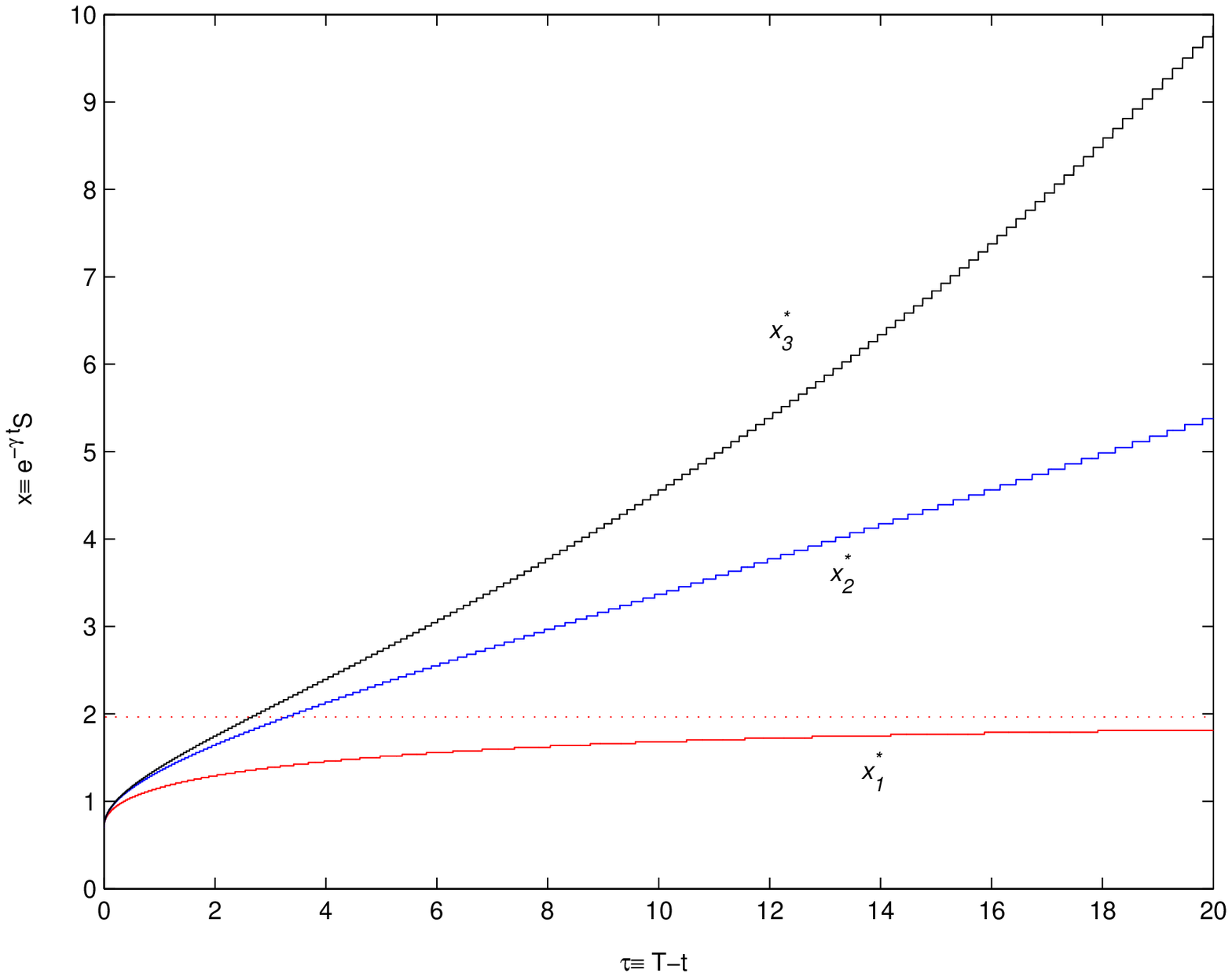,height=2.4in}%
\end{array}
\]
Figure 1: Optimal redeeming boundaries (Parameters: $r-\gamma =-0.04$, $%
\delta =0.03,$ $\sigma =0.4$ and $K=0.7.$)

\[
\begin{array}{c}
\psfig{figure=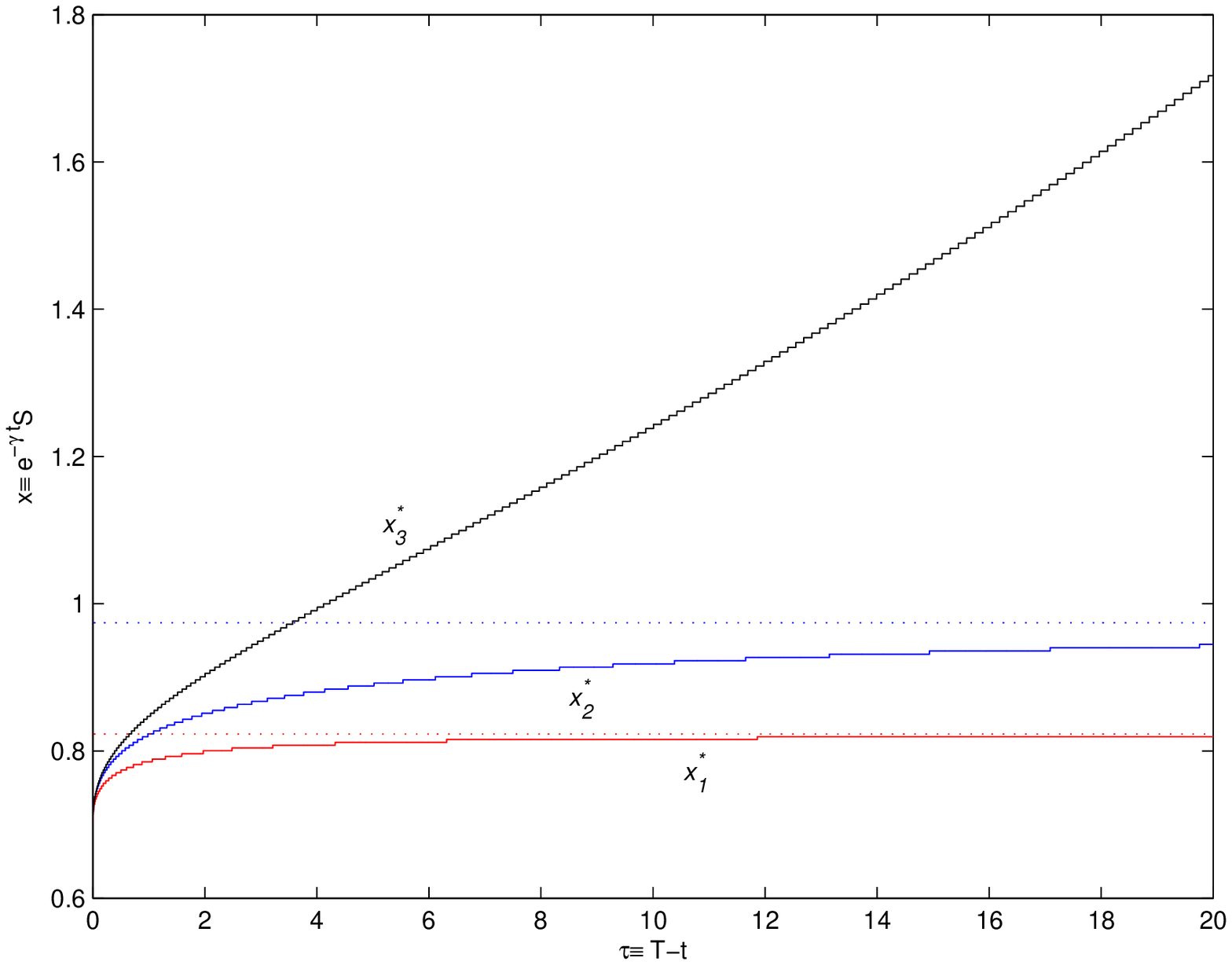,height=2.4in}%
\end{array}
\]
Figure 2: Optimal redeeming boundaries (Parameters: $r-\gamma =0.06$, $%
\delta =0.03,$ $\sigma =0.15$ and $K=0.7.$)

\ \newline
Now let us move on to the case in Section 5, where the pricing model,
resembling that of American-style Asian options, is a degenerate parabolic
variational inequality. We then employ the forward shooting grid method [cf.
Barraquand and Pudet (1996)], which is widely used to deal with this type of
degenerate problems. Figure 3 shows the optimal redeeming boundary $%
x_4^{*}(\cdot ,\cdot )$ in $x$-$A$-$\tau $ plane, where $x=e^{-\gamma t}S$, $%
A=e^{-\gamma t}I$ and $\tau =T-t.$ The data used are $r-\gamma =-0.04$, $%
\delta =0.03$, $\sigma =0.4$ and $K=0.7$. Time snapshots of the boundary are
depicted in Figure 4. It can be seen that the boundary, as a function of $A$
and $\tau ,$ is decreasing in $A,$ increasing in $\tau $ (i.e. decreasing in
$t$). At maturity the boundary is a straight line $x+A=K\equiv 0.7$.
Nevertheless, theoretical analysis indicates that $x_4^{*}(\cdot ,\tau )\leq
x_2^{*}(\tau )$. Figure 1 shows that approximately $x_2^{*}$ equals $1.35$
when $\tau =1,$ and $1.9$ when $\tau =3.$ It can be observed from Figure 4
that $x_4^{*}$ is indeed bounded from above by $x_2^{*}\left( \cdot \right) $%
. All these verify Proposition \ref{main1}.

\[
\begin{array}{c}
\psfig{figure=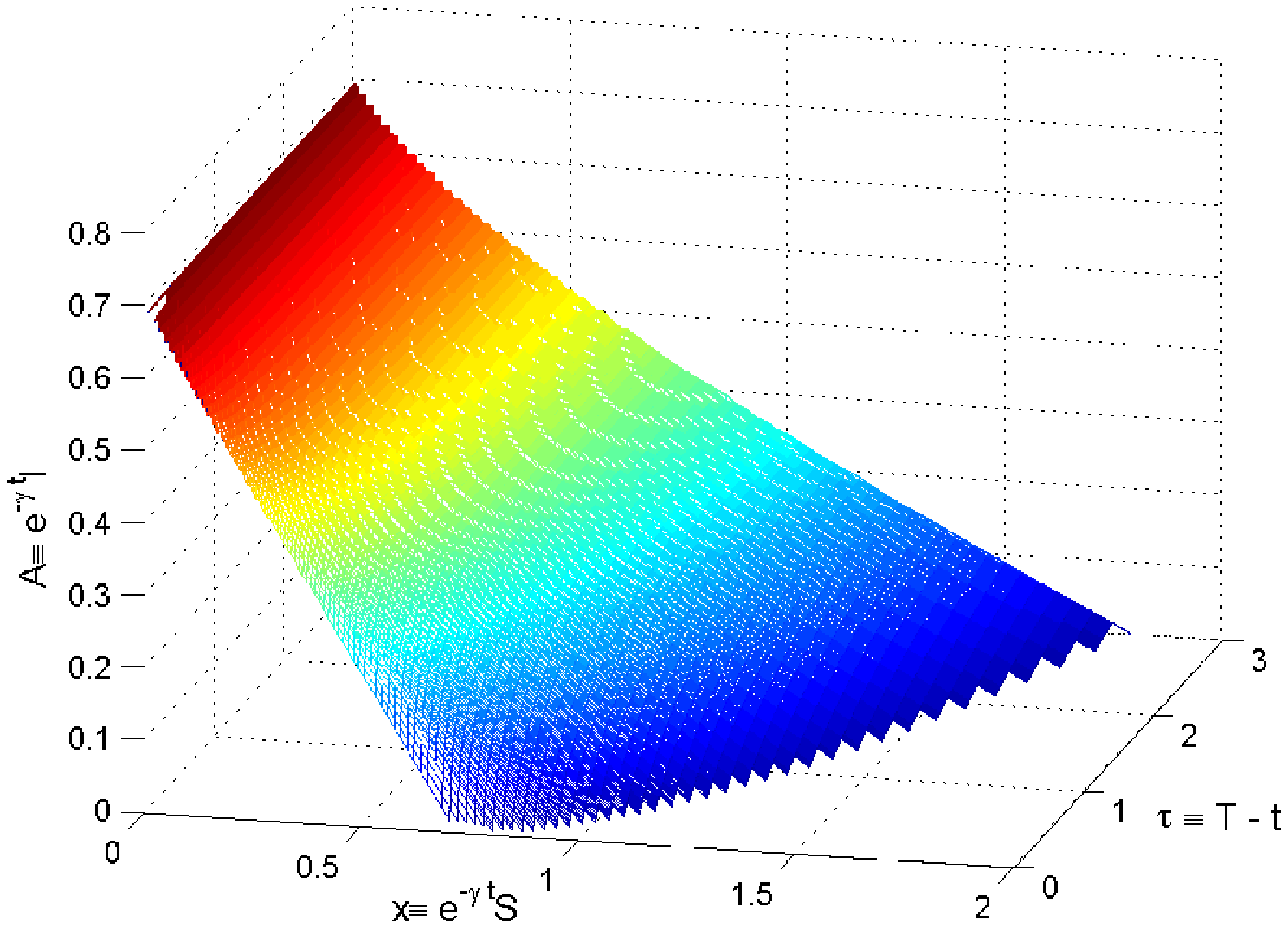,height=2.4in}%
\end{array}
\]
Figure 3: Optimal redeeming boundary $x_4^{*}(A,\tau )\equiv e^{-\gamma
t}S_4^{*}(I,t)$ (Parameters: $r-\gamma =-0.04$, $\delta =0.03$, $\sigma =0.4$
and $K=0.7$)

\[
\begin{array}{c}
\psfig{figure=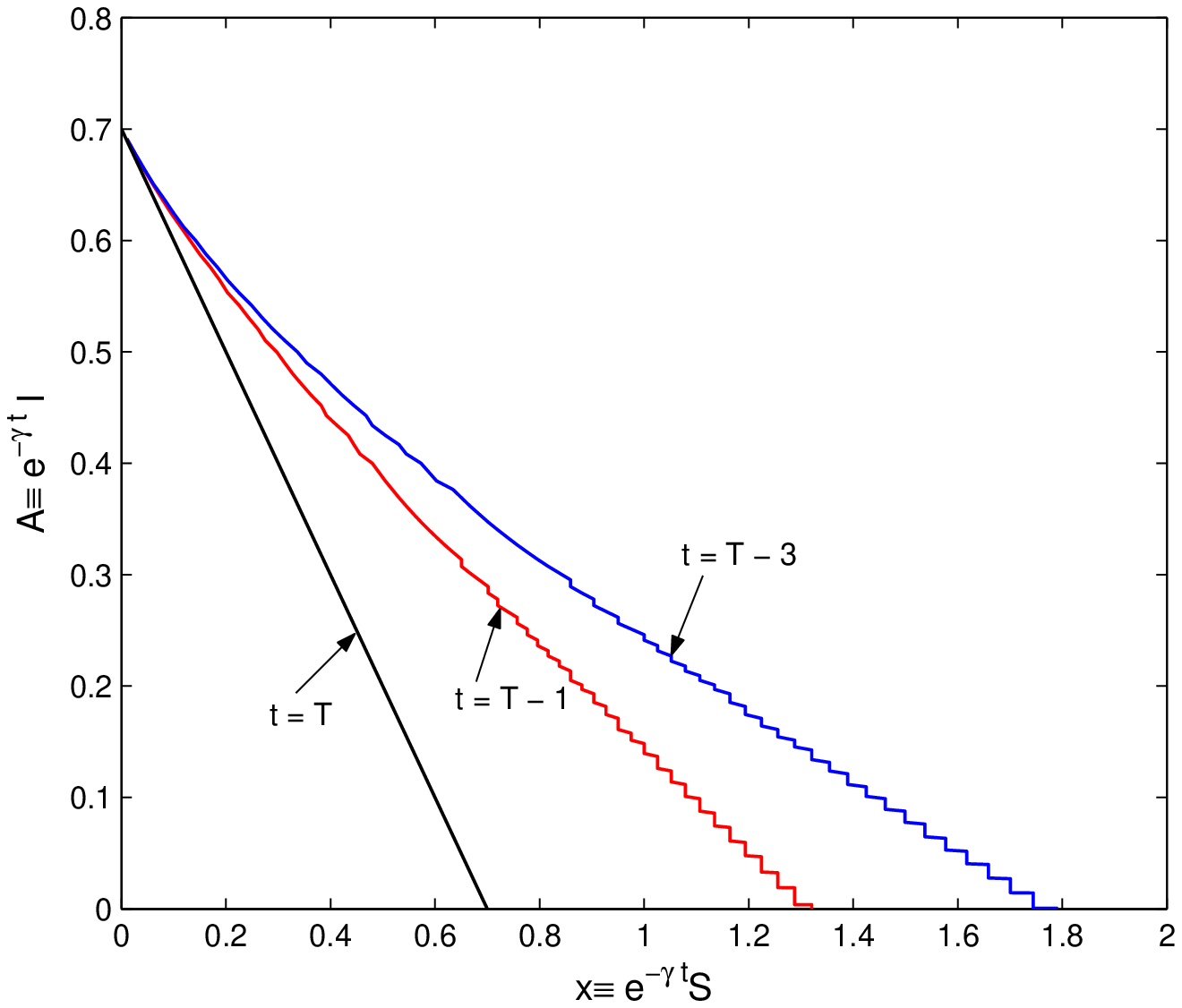,height=2.4in}%
\end{array}
\]
Figure 4: Time snapshots of $x_4^{*}(\cdot ,\cdot )$ (Parameters: $r-\gamma
=-0.04$, $\delta =0.03$, $\sigma =0.4$ and $K=0.7$)

\section{Conclusion and extensions}

\setcounter{equation}{0} \setcounter{theorem}{0}In the Black-Schole
framework, we formulate the pricing of stock loans as an optimal stopping
problem, or equivalently, a variational inequality. Since closed-form price
formulas are generally not available, we provide an analytic approach to
analyze the optimal redeeming strategy. It turns out that the way of
dividends distribution significantly alters the pricing model and the
optimal redeeming strategy. Numerical results are presented as well.

To conclude the paper, we briefly discuss some possible extensions of the
pricing models. Without loss of generality, we assume that the dividends are
gained by the lender and let $V=V(S,t)$ be the price function of the stock
loan.

\subsection{Amortized loan}

Assume that the loan is amortized. Let $C$ be the amortized rate defined as
follows:
\[
\int_{0}^{T}Ce^{\gamma (T-t)}dt=Ke^{\gamma T},
\]%
which yields $C=\frac{\gamma }{1-e^{-\gamma T}}K.$ At time $t,$ the amount
to be repaid is
\[
Ke^{\gamma t}-\int_{0}^{t}Ce^{\gamma (t-u)}du=\frac{C}{\gamma }\left(
1-e^{-\gamma (T-t)}\right) .
\]%
So, the early redemption payoff is $S_{t}-\frac{C}{\gamma }\left(
1-e^{-\gamma (T-t)}\right) ,$ $t\in \lbrack 0,T)$. Then the pricing model
becomes
\[
\left\{
\begin{array}{l}
\min \left\{ -L_{S}^{r,\delta }V+C,V-\left[ S-\frac{C}{\gamma }\left(
1-e^{-\gamma (T-t)}\right) \right] \right\} =0,\text{ } \\
V(S,T)=S,\hspace{1in}(S,t)\in Q.%
\end{array}%
\right.
\]

\subsection{Withdrawal feature}

The withdrawal feature aims to protect the lender from the extremely
downside risk of the collateralized stock. It allows the lender to
withdrawal the loan at any time by requesting the borrower to redeem the
stock with the price $L,$ where $L<K.$ This implies
\[
V(S,t)\leq L.
\]%
Like the pricing of game options, the model is described as a double
obstacle problem [cf. Kiefr (2000), Dai and Kwok (2005)] :%
\[
\left\{
\begin{array}{l}
\max \left\{ \min \left\{ -L_{S}^{r,\delta }V,V-\left( S-Ke^{\gamma
t}\right) \right\} ,V-L\right\} =0,\text{ } \\
V(S,T)=\min \left\{ (S-Ke^{\gamma T})^{+},L\right\} ,\hspace{1in}(S,t)\in Q.%
\end{array}%
\right.
\]

\subsection{Renewal feature}

If we assume that the borrower has right to renew the loan, this will lead
to multiple stopping problems. See, for example, Carmona and Touzi (2003),
IBanez (2004), Meinshausen and Hambly (2004) and Dai and Kwok (2008). From
the point of view of PDEs, the pricing model is described by a series of
variational inequalities [cf. Dai and Kwok (2008)]. We refer interested
readers to the references.

\vspace{1.5cm}
\renewcommand{\theequation}{A-\arabic{equation}}
\renewcommand
\thesection{Appendix:}
\setcounter{equation}{0} 
\appendix

\begin{center}
\textbf{Appendix: The proof of Proposition \ref{xz}}
\end{center}

Applying the maximum principle, it is easy to see that the solution to
problem (\ref{eq1}) possesses the following properties:
\begin{equation}
0\leq f(x,t)\leq 1\text{ and }0\leq \frac{\partial f}{\partial x}\leq 1.
\label{cond1}
\end{equation}

Denote $f_{\infty }(x)=\lim_{\left( T-t\right) \rightarrow +\infty }f(x,t).$
Then $f_{\infty }(x)$ satisfies the stationary counterpart of (\ref{eq1}),
which is equivalent to a free boundary problem:
\begin{eqnarray}
-\frac{1}{2}\sigma ^{2}x^{2}\frac{\partial ^{2}f_{\infty }}{\partial x^{2}}%
-\left( \overline{r}-\delta \right) x\frac{\partial f_{\infty }}{\partial x}+%
\overline{r}f_{\infty } &=&0,\text{ }x\leq x_{\infty }^{\ast }  \label{fb1}
\\
f_{\infty }(x_{\infty }^{\ast }) &=&x-K,\text{ and }f_{\infty }^{^{\prime
}}(x_{\infty }^{\ast })=1.  \label{fb3}
\end{eqnarray}%
Here $x_{\infty }^{\ast }$ is the free boundary to be determined. Due to (%
\ref{cond1}), we are only concerned with the solution to (\ref{fb1})-(\ref%
{fb3}) under the restrictions
\begin{equation}
0\leq f_{\infty }(x)\leq x\text{ and }0\leq f_{\infty }^{^{\prime }}(x)\leq
1.  \label{cond2}
\end{equation}

As is well-known, the general solution to equation (\ref{fb1}) is
\[
f_\infty (x)=C_1x^{\alpha _{+}}+C_2x^{\alpha _{-}},
\]
where $C_1$ and $C_2$ are to be determined, and $\alpha _{+}$ (as given in (%
\ref{alpha})) and $\alpha _{-}=1-\frac{2(r-\gamma -\delta )}{\sigma ^2}%
-\alpha _{+}$ are two roots of the algebraic equation $\frac{\sigma ^2}%
2\alpha ^2+(\overline{r}-\delta -\frac{\sigma ^2}2)\alpha -\overline{r}=0.$
Let us first assume $\delta >0.$ It is easy to check that $\alpha
_{-}<1<\alpha _{+},$ which yields $C_2=0$ or
\begin{equation}
f_\infty (x)=C_1x^{\alpha _{+}}.  \label{fsol}
\end{equation}
Otherwise we would have $f_\infty ^{^{\prime }}(x)=C_1\alpha _1x^{\alpha
_{+}-1}+C_2\alpha _2x^{\alpha _{-}-1}\rightarrow \infty $ as $x\rightarrow
0, $ which contradicts (\ref{cond2}).

From (\ref{fsol}), we can make use of (\ref{fb3}) to get
\begin{equation}
C_{1}=\frac{1}{\alpha _{+}}\left( \frac{\alpha _{+}-1}{\alpha _{+}K}\right)
^{\alpha _{1}-1}\text{ and }x_{\infty }^{\ast }=\frac{\alpha _{+}}{\alpha
_{+}-1}K  \label{eqx}
\end{equation}%
for $\delta >0.$ To deal with the case of $\delta =0,$ we let $\delta $ go
to $0$ in (\ref{eqx})$.$ It is easy to see that if $r+\frac{1}{2}\sigma
^{2}\geq 0$ and $\delta \rightarrow 0,$ then $\alpha _{+}\rightarrow 1,$
which yields $x_{\infty }^{\ast }=+\infty .$ If $\overline{r}+\frac{1}{2}%
\sigma ^{2}<0$ and $\delta \rightarrow 0,$ then $\alpha _{+}>1$ and the
expression of $x_{\infty }^{\ast }$ is the same as in (\ref{eqx}). The proof
is complete.\qed

\end{document}